\documentclass[12pt,prd,aps,amssymb,amsmath,tightenlines,showkeys]{revtex4}

\usepackage{graphicx}

\newcommand{\txt}{\textstyle}
\newcommand{\dsp}{\displaystyle}
\newcommand{\beq}{\begin{equation}}
	\newcommand{\eeq}{\end{equation}}
\newcommand{\bea}{\begin{eqnarray}}
	\newcommand{\eea}{\end{eqnarray}}
\newcommand{\ba}{\begin{array}}
	\newcommand{\ea}{\end{array}}

\newcommand{\half} {{\txt \frac{1}{2}}}

\newcommand{\e}{{\rm e}}   

\newcommand{\rd}{{\rm d}}

\newcommand{\nn}{\nonumber}

\newcommand{\eg}{{e.g.}}

\begin{document}
	
\title{Exactly solvable nonlinear eigenvalue problems}
\author{Qing-hai Wang}
\email{qhwang@nus.edu.sg}
\affiliation{Department of Physics, National University of Singapore, 117551, Singapore}
	
\date{24 July 2020}
	
\begin{abstract}
The nonlinear eigenvalue problem of a class of second order semi-transcendental differential equations is studied. A nonlinear eigenvalue is defined as the initial condition which gives rise a separatrix solution. A semi-transcendental equation can be integrated once to a first order nonlinear equation, \eg, the Ricatti equation. It is shown that the nonlinear eigenvalue problems of these semi-transcendental equations are equivalent to linear eigenvalue problems. They share the exactly same eigenvalues. The eigensolutions in the two problems are closely related. The nonlinear eigenvalue problem equivalent to the (half) harmonic oscillator in quantum mechanics is solved exactly. This is the first solvable nonlinear eigenvalue problem. The nonlinear eigenvalue problems of some extended equations are also studied.	
\end{abstract}
	
\keywords{Riccati equation, semiclassical analysis, asymptotic approximation, eigenvalue}
	
\maketitle
\section{Introduction}
\label{sec:intro}
The nonlinear eigenvalue problem was first proposed in 2014 by Bender \textit{et al} as an initial-value problem for the first order nonlinear differential equation \cite{Bender1}
\begin{equation}
y'(x)= \cos[\pi x y(x)].
\end{equation}
For large $x$, the leading asymptotic behaviour for the solution of the above equation is
\begin{equation}
y(x)\sim \left(m+\half\right)/x, \quad x\to\infty,
\end{equation}
where $m$ is an integer. When $m=2n$ is even, the asymptotic behaviour is stable. That is, if the initial condition $y(0)=a$ gives a solution approaching to $y(x)\sim \left(2n+\half\right)/x$, then a slightly different initial condition $y(0)=a+\delta$ will give a solution with the same asymptotic behaviour when $\delta$ is sufficiently small. Thus, solutions with slightly different initial conditions bundle together for large $x$. On the contrary, when $m=2n+1$ is odd, the asymptotic behaviour is unstable. Only a precise initial condition, say $y(0)=b_n$, can give rise a solution with $y_n(x)\sim \left(2n+1+\half\right)/x$. Any small deviation in the initial condition, $y(0)=b_n+\delta$, will lead the solution to either $y(x)\sim \left(2n+\half\right)/x$ or  $y(x)\sim \left[2(n+1)+\half\right]/x$. This is true regardless how small $\delta$ is, as long as $\delta\neq0$. Therefore, the solution $y_n(x)$ is \textit{a separatrix solution}. This unstable feature is somewhat similar to the solutions in linear eigenvalue problems near an eigensolution. In a typical linear eigenvalue problem, \eg, the harmonic oscillator, an exact eigenvalue gives rise an eigensolution which decays as $x\to\infty$ and $x\to-\infty$. A slight deviation from the exact eigenvalue will not lead to a solution which is slightly different from the eigensolution. Rather, its magnitude blows up as $x\to\infty$ or $x\to-\infty$. In this sense, we call $y_n(x)$ as \textit{an eigensolution}, and the corresponding initial conditions $b_n$ as \textit{the nonlinear eigenvalue}.  

Later, the nonlinear eigenvalue problem was extended to the first two Painlev\'e equations \cite{Bender1,Bender2}: 
\begin{eqnarray}
y''(x)&=& 6 y^2(x)+x, \label{eqn:PI}\\
y''(x)&=& 2 y^3(x) + y(x). 
\label{eqn:PII}
\end{eqnarray}
For these second order differential equations, two types of nonlinear eigenvalue problems were studied. One is holding the initial function $y(0)$ fixed and varying the initial slope $y'_n(0)$ as the eigenvalues. Another one is varying $y_n(0)$ with a fixed $y'(0)$. For large $n$, the nonlinear eigenvalues follow power laws in the leading asymptotic behaviour. These laws were first found numerically \cite{Bender1,Bender2}. They were then proved rigorously \cite{Liu,Long2}. Curiously, these power laws are deeply connected to the linear eigenvalues of some non-Hermitian Hamiltonians in the WKB approximation \cite{BB98}.

Recently, the first two Painlev\'e equations in Eqs.~(\ref{eqn:PI}) and (\ref{eqn:PII}) were generalised to an infinite series equations with the movable singularities being algebraic \cite{Bender3,Bender4}
\begin{equation}
y''(x)=\textstyle{\frac{2(M+1)}{(M-1)^2}}\left[y^M(x)+xp(y)+q(y)\right],
\label{eqn:GP}
\end{equation}
where $p(y)$ and $q(y)$ are polynomials in $y$ of degree at most $M-2$:
\begin{eqnarray}
p(y)&=& P_{M-2}\,y^{M-2}+P_{M-3}\,y^{M-3}+\cdots+P_1\,y+P_0,\nonumber\\
q(y)&=& Q_{M-2}\,y^{M-2}+Q_{M-3}\,y^{M-3}+\cdots+Q_1\,y+Q_0.\nonumber
\end{eqnarray}
Clearly that the first Painlev\'e equation in Eq.~(\ref{eqn:PI}) is the special case with $M=2$ and the second Painlev\'e equation in Eq.~(\ref{eqn:PII}) is the special case with $M=3$. 

Numerical nonlinear eigenvalues and eigensolutions were obtained for some generalised Painlev\'e equations. The asymptotic behaviours of large nonlinear eigenvalues exhibit similar power laws as in the first two Painlev\'e equations \cite{Bender3,Bender4}. A new type of nonlinear eigenvalue problem was discovered, which resembles to some characteristic features of the hyperfine structure in nuclear physics \cite{Bender4}.      

In this paper, we study the nonlinear eigenvalue problem of another second order differential equation with the form
\begin{equation}
\frac{\rd^2W(x)}{\rd x^2} = -2W(x) \frac{\rd W(x)}{\rd x} + q(x) \frac{\rd W(x)}{\rd x} + q'(x) W(x),
\label{eqn:InceV}
\end{equation}
where $q(x)$ is an arbitrary analytic function and $q'(x)$ is its derivative. This equation is called \textit{semi-transcendental} because it can be integrated once to a first order nonlinear equation. In this particular example, the first order equation is equivalent to the Ricatti equation. In Sec.~\ref{sec:Ric}, we show an equivalent relation between the nonlinear eigenvalue problem of Eq.~(\ref{eqn:InceV}) and a linear eigenvalue problem based on this semi-transcendental feature. Unlike all the \textit{asymptotic} results on the nonlinear eigenvalue problems studied so far, this equivalent relation is \textit{exact}. This is the primary result of this paper. It puts the nonlinear eigenvalue problem on a much solid ground.

For some special choices of $q(x)$ in Eq.~(\ref{eqn:InceV}), the resulting Riccati equation is exactly solvable. To illustrate the equivalent relation, we present two examples in Sec.~\ref{sec:exact}. The first one is an exactly solvable nonlinear eigenvalue problem which is equivalent to a half simple harmonic oscillator. The nonlinear eigenvalues and eigensolutions are solved exactly. To our best knowledge, this is the first exactly solvable nonlinear eigenvalue problem. The second example is a numerical study on the nonlinear eigenvalue problem which is equivalent to a half quartic anharmonic oscillator.

All the movable singularities of first two Painlev\'e equations in Eqs.~(\ref{eqn:PI}) and (\ref{eqn:PII}) are poles. They can be generalised into equations in (\ref{eqn:GP}) with movable singularities being algebraic. Similarly, we generalise Eq.~(\ref{eqn:InceV}) into a series of equations with algebraic movable singularities in Sec.~\ref{sec:ext}. The nonlinear eigenvalues of two such equations are obtained numerically. The asymptotic behaviour of large eigenvalues are found to follow simple power laws, just like the first two Painlev\'e equations and the generalised Painlev\'e equations.

\section{Equivalence of two types of eigenvalue problems}
\label{sec:Ric}

In this section we present an equivalent relation between a nonlinear eigenvalue problem and a linear eigenvalue problem. By exploiting the semi-transcendental feature of Equation (\ref{eqn:InceV}), we show that the two eigenvalue problems possess essentially same eigenvalues, up to a minus sign and a constant shift. The eigensolutions in the two problems are closely related. 

Equation (\ref{eqn:InceV}) is the number V equation of the Canonical Equations of Type I in the classic textbook on ordinary differential equations by Ince \cite{Ince}. It can obviously be integrated once to a first-order nonlinear equation, which is equivalent to the Ricatti equation.  To see this, let us first change the dependent variable as
\begin{equation}
y(x)=W(x)-\half q(x).
\end{equation}
The equation satisfied by $y(x)$ is 
\begin{equation}
y''(x)=-2y'(x)y(x)+V'(x),
\label{eqn:master}
\end{equation}
where we have defined the derivative of an arbitrary function $V(x)$ as
\begin{equation}
V'(x)\equiv \half q'(x)q(x)-\half q''(x).
\end{equation}
For any analytic function $V(x)$, the movable singularities of $y(x)$ are all simple poles. 

Equation (\ref{eqn:master}) can be directly integrated to a Riccati equation,
\begin{equation}
y'(x)=-y^2(x) +V(x) - E,
\label{eqn:Recatti}
\end{equation}
where $E$ is a constant of integration.

If $V(x)$ is unbounded for large $x$, then $y(x)$ may either develop infinite movable singularities, or, after finite singular points, approach to one of the two asymptotes
\begin{equation}
y(x)\sim\pm\sqrt{V(x)- E}, \quad x\to\infty.
\end{equation}
It can be shown that for positive $x$, the upper branch is a stable fixed point in the functional space and the lower branch is unstable. To see this, let 
\begin{equation}
y(x)=Y(x)+\phi(x),
\end{equation}
where $Y(x)$ is a solution of Eq.~(\ref{eqn:Recatti}). Plugging into Eq.~(\ref{eqn:Recatti}), we get
\begin{equation}
\phi'(x)=-2Y(x)\phi(x)-\phi^2(x).
\end{equation}
The solution to this nonlinear equation is
\begin{equation}
	\phi(x) = \frac{\e^{-2\int_a^x\rd s\, Y(s)}}{\dsp \int_b^x\rd t \,\e^{-2\int_a^t\rd s\, Y(s)}},
\label{eqn:IbP}	
\end{equation}
where $a$ and $b$ are two constants. Note that $\phi(x)$ depends only on $b$, but not on $a$. Depending on the sign of $Y(x)$, the denominator of Eq.~(\ref{eqn:IbP}) has two distinct asymptotic behaviours for large $x$ \cite{BenderBook}:
\begin{equation}
	\int_b^x\rd t \,\e^{-2\int_a^t\rd s\, Y(s)} \sim
	\left\{
		\begin{array}{lll}
			\dsp \int_b^\infty\rd t \,\e^{-2\int_a^t\rd s\, Y(s)} \equiv C, & Y(x) \to +\infty, & x \to \infty; \\
			-\dfrac{\e^{-2\int_a^x\rd s\, Y(s)}}{2Y(x)}, & Y(x)\to -\infty, & x \to \infty.
		\end{array}
	\right.
\end{equation}
This in turn leads two asymptotic behaviours for $\phi(x)$,
\begin{equation}
	\phi(x)\sim 
	\left\{
		\begin{array}{lll}
		\dsp \frac{1}{C} \e^{-2\int_a^x\rd s\, Y(s)}\to 0, & Y(x) \to +\infty, & x \to \infty; \\
		-2Y(x)\to\infty, & Y(x)\to -\infty, & x \to \infty.
		\end{array}
	\right.
\label{eqn:critical}
\end{equation}
Thus, the positive branch $Y(x)\sim \sqrt{V(x) - E}$ is stable and the negative branch is unstable. Note that $\phi(x)$  does not change signs for a real solution $Y(x)$. This means that, near the stable fixed point (asymptote), the solutions approach it without oscillation. This is different from the first two Painlev\'e equations and their generalisations studied in Refs.~\cite{Bender1,Bender2,Bender3,Bender4}.

A Riccati equation is equivalent to a second order linear differential equation. To see this,  let us rewrite the solution of Eq.~(\ref{eqn:Recatti}) as
\begin{equation}
y(x)=\frac{\psi'(x)}{\psi(x)},
\end{equation}  
then $\psi(x)$ satisfies a Schr\"odinger equation
\begin{equation}
-\psi''(x)+V(x)\psi(x)=E\psi(x).
\label{eqn:Schr}
\end{equation}

The solution to the Schr\"odinger equation (\ref{eqn:Schr}) has two asymptotic behaviors for large $x$:
\begin{equation}
\psi(x)\sim \frac{N_\pm}{[V(x)-E]^{1/4}} \e^{\pm\int_0^x\rd s\sqrt{V(s)-E}}, \quad x\to\infty.
\end{equation}
Here we have chosen the lower integration limit to be $0$ without loss of generality. Only the diminishing solutions are corresponding to the separatrix solutions in $y(x)\sim-\sqrt{V(x)-E}$. That is, $N_+=0$. 

For the initial slope nonlinear eigenvalue problem of Eq.~(\ref{eqn:master}), the initial value $y(0)$ is fixed and the initial slope $y'(0)$ is the eigenvalue when it yields a separatrix solution. If we choose a homogeneous initial value as  
\begin{equation}
y(0)=0,
\end{equation}
then the initial slope of $\psi$ vanishes, 
\begin{equation}
\psi'(0)=0.
\end{equation}
That is, to find a separatrix solution of $y$ is equivalent to solve the Schr\"odinger equation with the boundary conditions $\psi(+\infty)=0$ and $\psi'(0)=0$. The eigenvalue of the \textit{linear} eigenvalue problem is simply $E$. The eigenvalue of the initial slope problem of the \textit{nonlinear} eigenvalue problem is 
\begin{equation}
y'(0)=V(0)-E.
\end{equation}
We emphasise that this equivalence between the two eigenvalue problems is exact.

\section{An exactly solvable model}
\label{sec:exact}
The nonlinear differential equation (\ref{eqn:master}) can be solved exactly with a quadratic function $V$,
\begin{equation}
V(x)=x^2.
\end{equation}
The corresponding Schr\"odinger equation (\ref{eqn:Schr}) describe the quantum harmonic oscillator. 

A separatrix solution has the form 
\begin{equation}
y(x)=\frac{D'_{\nu}(x)}{D_{\nu}(x)}, 
\end{equation}
where $D_\nu(x)$ is the parabolic cylinder function and $\nu$ is related to the integration constant in Eq.~(\ref{eqn:Recatti}) by
\begin{equation}
E=2\nu+1.
\end{equation}
At the origin, we have the initial conditions
\begin{equation}
y(0) = -2\frac{\Gamma\left(-\frac{\nu}{2}-\half\right)}{\Gamma\left(-\frac{\nu}{2}\right)}, \qquad 
y'(0) =-2\nu-1 -4\frac{\Gamma^2\left(-\frac{\nu}{2}-\half\right)}{\Gamma^2\left(-\frac{\nu}{2}\right)}.
\end{equation}
For real parameter $\nu$, the above $y'(0)$ never vanishes. Therefore, there is no nonlinear eigenvalues for the initial function problem with $y'(0)=0$. 

For the initial slope problem with $y(0)=0$, we get 
\begin{equation}
\nu=2n, \qquad n=0,1,2,\cdots.
\end{equation} 
The eigenfunctions are
\begin{equation}
y_n(x)=-x + 4n \frac{H_{2n-1}(x)}{H_{2n}(x)}, \qquad n=0,1,2,\cdots.
\end{equation}
The nonlinear eigenvalues are
\begin{equation}
y_n'(0)=-4n-1.
\end{equation}

Similarly, if one chooses 
\begin{equation}
V(x)=x^4,
\end{equation}
half of the eigenvalues of the anharmonic oscillator will be the eigenvalues of the nonlinear initial slope problem (see Table \ref{tab:anHO}). There is no real eigenvalues for the initial function problem for this equation with $y'(0)=0$.
\begin{center}
	\begin{table}[htb]
		\begin{center}
		\caption{\label{tab:anHO}
	Initial slopes which give rise to separatrix solutions to the equation $y''(x)=-2y'(x)y(x)+4x^3$ with $y(0)=0$.}
			\begin{tabular}{cr}
				\hline
				\hline
				$n$ & $y'_n(0)\qquad$\\
				\hline
				 0& $-1.06036209$\\
				 1& $-7.45569793$\\
				 2&$-16.26182601$\\
				 3&$-26.52847118$\\
				 4&$-37.92300102$\\
				 5&$-50.25625451$\\
				 6&$-63.40304698$\\
				 7&$-77.27320048$\\
				 8&$-91.79806681$\\
				\hline
				\hline
			\end{tabular}
		\end{center}
	\end{table}
\end{center}

\section{Extended Ricatti equations}
\label{sec:ext}

In this section, we generalise the (derivative) of the Riccati equation in Eq.~(\ref{eqn:InceV}) to a class of semi-integrable equations with movable singularities to be algebraic. Namely, there is no logarithmic or other essential singularities. This is a similar work which has been done to the first two Painlev\'e equations in \cite{Bender3,Bender4}. Since the term ``generalised Riccati equations'' was used for other purpose, we call these new equations as ``extended Ricatti equations.'' The generic form for these equations are
\begin{equation}
y''(x) = - \frac{n+1}{n} y'(x) y^n(x) + \sum_{i=0}^{m} \sum_{j=0}^{k} a_{mk}x^m y^k(x).
\end{equation}

For example, for the nonlinear equation
\begin{equation}
y''(x)=-\frac{4}{3}y'(x)y^3(x)+x,
\end{equation}
the movable singularities are cubic roots:
\begin{eqnarray}
y(x) &\sim& \frac{1}{(x-x_0)^{1/3}} + a_3 (x-x_0)^1 + \frac{3x_0}{10} (x-x_0)^{2}\\ \nn 
&&\quad  - \frac{6a_3^2}{11} (x-x_0)^{7/3} + \frac{3}{26} (x-x_0)^{3} + \sum_{i=10}^\infty a_i (x-x_0)^{i/3}, \qquad x\to x_0,
\end{eqnarray}
where $a_3$ is a constant to be determined by the initial conditions. 

For $y(0)=0$, the nonlinear eigenvalues for the initial slope problem are listed in Table \ref{tab:GR3a}. For large $n$, they follow a simple power law
\begin{equation}
y'_n(0)\sim -2.32805 n^{4/5},\qquad n\to\infty.
\end{equation}

\begin{center}
	\begin{table}[htb]
		\begin{center}
		\caption{	\label{tab:GR3a}
			Initial slopes which give rise to separatrix solutions to the equation $y''(x)=-\frac{4}{3}y'(x)y^3(x)+x$ with $y(0)=0$.}
			\begin{tabular}{rr}
				\hline
				\hline
				$n$ & $y'_n(0)\qquad$\\
				\hline
				0& $ -1.00243383$ \\
				1& $ -2.94953545$ \\
				2& $ -4.60069679$\\
				3& $ -6.11303401$\\
				4& $ -7.53640737$\\
				5& $ -8.89523799$\\
				6& $-10.20390084$\\
				7& $-11.47180488$\\
				8& $-12.70555477$\\
				9& $-13.91002352$\\
				10&$-15.08894332$\\
				11&$-16.24525712$\\
				12&$-17.38134027$\\
				13&$-18.49914706$\\
				14&$-19.60031125$\\
				15&$-20.68621727$\\
				16&$-21.75805200$\\
				17&$-22.81684331$\\
				18&$-23.86348925$\\
				19&$-24.89878065$\\
				20&$-25.92341874$\\
				21&$-26.93802924$\\
				22&$-27.94317361$\\
				23&$-28.93935824$\\
				24&$-29.92704197$\\
				25&$-30.90664231$\\
				\hline
				\hline
			\end{tabular}
		\end{center}
	\end{table}
\end{center}

Similarly, for the equation
\begin{equation}
y''(x)=-\frac{4}{3}y'(x)y^3(x)+x y(x),
\end{equation}
the movable singularities are also cubic roots:
\begin{eqnarray}
y(x)&\sim& \frac{1}{(x-x_0)^{1/3}} +a_3 (x-x_0)^1 + \frac{x_0}{2} (x-x_0)^{5/3} - \frac{6a_3^2}{11} (x-x_0)^{7/3}\\ \nn 
&&\quad + \frac{3}{20} (x-x_0)^{8/3} - \frac{9 a_3 x_0}{26} (x-x_0)^{3} + \sum_{i=10}^\infty a_i (x-x_0)^{i/3}, \qquad x\to x_0.
\end{eqnarray}
For $y(0)=0$, the nonlinear eigenvalues for the initial slope problem are listed in Table \ref{tab:GR3b}. For large $n$, they follow a simple power law
\begin{equation}
y'_n(0)\sim -0.96951 n^{4/9},\qquad n\to\infty.
\end{equation}

\begin{center}
	\begin{table}[htb]
		\begin{center}
			\caption{	\label{tab:GR3b}
				Initial slopes which give rise to separatrix solutions to the equation $y''(x)=-\frac{4}{3}y'(x)y^3(x)+x y(x)$ with $y(0)=0$.}
			\begin{tabular}{rr}
				\hline
				\hline
				$n$ & $y'_n(0)\qquad$\\
				\hline
				0& $-0.89134081$ \\
				1& $-1.21797331$ \\
				2& $-1.48616058$\\
				3& $-1.71029522$\\
				4& $-1.90473552$\\
				5& $-2.07799763$\\
				6& $-2.23533407$\\
				7& $-2.38019349$\\
				8& $-2.51496483$\\
				9& $-2.64137602$\\
				10&$-2.76072118$\\
				11&$-2.87399742$\\
				12&$-2.98199128$\\
				13&$-3.08533557$\\
				14&$-3.18454800$\\
				15&$-3.28005827$\\
				\hline
				\hline
			\end{tabular}
		\end{center}
	\end{table}
\end{center}

\section{Conclusions}
\label{sec:concl}

In this paper, we present an exact equivalence between a class of nonlinear eigenvalue problems and conventional linear eigenvalue problems. By exploiting the semi-integrable features of these nonlinear differential equations, we have linked the nonlinear eigenvalues to the well-known linear eigenvalues. We found the first exactly solvable nonlinear eigenvalue problem. This equivalence also provides new insights about the nature of nonlinear eigenvalue problems. 

In addition, we generalised the Riccatti equations to an infinite series of nonlinear equations whose movable singularities are algebraic. Two examples are explicitly presented. The nonlinear eigenvalues exhibit a similar power law as those in the Painlev\'e equations. There are totally 50 second order differential equations satisfying the Painlev\'e criterion \cite{Ince}. Many of them may be generalised to have algebraic movable singularities. We leave this to a future study.    

\begin{acknowledgments}
	The author thanks for the useful discussions with Carl M.~Bender. The author is grateful for the comments and suggestions on the writing from Jiangbin Gong.
\end{acknowledgments}

\end{document}